\documentclass[12pt, nofootinbib, amsfonts, amsmath, amssymb, a4paper, prd, preprint]{revtex4}

\usepackage{bbold}
\usepackage{braket}
\usepackage{color}

\newcommand{\del}{\ensuremath{\partial}}
\newcommand{\ud}{\mathrm{d}}
\newcommand{\mJ}{\mathcal{J}}
\newcommand{\mK}{\mathcal{K}}
\newcommand{\mW}{\mathcal{W}}

\begin{document}

\title{Vertex functions and their flow equations from the\\ 2PI effective action}
\author{Peter Millington}
\email{peter.millington@manchester.ac.uk}
\affiliation{Department of Physics and Astronomy, University of Manchester,\\ Manchester M13 9PL, United Kingdom}
\author{Paul M.\ Saffin}
\email{paul.saffin@nottingham.ac.uk}
\affiliation{School of Physics and Astronomy, University of Nottingham,\\ Nottingham NG7 2RD, United Kingdom}
\date{8 November 2022}

\begin{abstract}

    By exploiting the convexity of the two-particle-irreducible (2PI) effective action, we describe a procedure for extracting $n$-point vertex functions.  This procedure is developed within the context of a zero-dimensional ``quantum field theory'' and subsequently extended to higher dimensions. These results extend the practicability and utility of a recent, alternative approach to the functional renormalization group programme [see Phys.\ Rev.\ D 104 (2021) 069906; J.\ Phys.\ A 54 (2021) 465401], and clarify the relationship between the flow equations for coupling parameters and vertices.
    
    \bigskip
    
    \begin{footnotesize}
    \noindent This is an author-prepared post-print of \href{https://doi.org/10.1088/1751-8121/ac99ae}{J.\ Phys.\ A:\ Math.\ Theor.\ {\bf 55} (2022) 435402}, published by IOP Publising under the terms of the \href{https://creativecommons.org/licenses/by/4.0/}{CC BY 4.0} license.
    \end{footnotesize}

\end{abstract}

\maketitle


\section{Introduction}

In a series of works, we have used a zero-dimensional ``quantum field theory'' to provide explicit expositions of a number of features of the two-particle irreducible (2PI) effective action.\footnote{The relativistic 2PI effective action was described by Coleman, Jackiw and Tomboulis~\cite{Cornwall:1974vz}, building on early applications of higher-order Legendre transforms in non-relativistic statistical mechanics, e.g., by De Dominicis and Martin~\cite{deDominicis:1964zz, deDominicis2}.}  This has included an exposition of the behaviours of the external sources in its definition (see also Ref.~\cite{Garbrecht:2015cla}), as well as the true convex-conjugate pairs of variables that correspond respectively to the one- and non-connected two-point functions~\cite{Millington:2019nkw}.\footnote{The importance of identifying the true convex-conjugate variables, in terms of the non-connected functions, was emphasised earlier in Ref.~\cite{Vasilev:1973vd}.} In the same work, we were able to visualise the convexity of the 2PI effective action, illustrate how non-convexity emerges for particular constraints on the two-point function, describe how the number and type of the saddle points is fixed by these constraints, and recover the Maxwell construction for a classical action with two local minima (and mirroring the analysis of Ref.~\cite{Rivers:1983sq}). Subsequently, by exploiting the convexity of the 2PI effective action, we were able to make use of the identities reported in footnote 11 of Coleman, Jackiw and Tomboulis~\cite{Cornwall:1974vz} to derive an expression for the two-point function in terms of partial derivatives of the 2PI effective action~\cite{Millington:2021ftp}. Similar expressions can be found in, e.g., Ref.~\cite{Berges:2005hc}, (cf.~the `external propagator' of Ref.~\cite{vanHees:2002bv}), and many of the functional identities reported in this work can also be found (again derived by means of convexity) in the much earlier and little-known series of works by Vasil'ev~\cite{Vasilev:1973vd}, and Vasil'ev and Kazanskii~\cite{Vasilev1, Vasilev2}.

The resulting expression for the two-point function allowed us to provide a detailed comparison of a new approach to deriving the exact flow equations of the functional renormalization group based on the 2PI effective action~\cite{Alexander:2019cgw} with the well-known approach based on the average one-particle-irreducible (1PI) effective action~\cite{Wetterich:1992yh, Morris:1993qb, Ellwanger:1993mw} (see also Ref.~\cite{Reuter:1996cp} in the context of gravity, and, e.g., Refs.~\cite{Bagnuls:2000ae, Berges:2000ew, Pawlowski:2005xe, Gies:2006wv, Kopietz:2010zz, Rosten:2010vm, Dupuis:2020fhh} for reviews).\footnote{Recently, zero-dimensional quantum field theories have also been used to study the functional renormalization group in the context of $O(N)$ models~\cite{Koenigstein:2021syz, Koenigstein:2021rxj, Steil:2021cbu}.} However, in order for this new approach to be of utility, it is necessary to formulate a procedure for extracting the $n$-point vertex functions from the 2PI effective action, and this is the focus of the present work. As we will see, this involves taking partial derivatives of the 2PI effective action with respect to both the one- and two-point functions, differing therefore from the corresponding procedure for the 1PI~\cite{Jackiw:1974cv} or average 1PI~\cite{Wetterich:1989xg} effective actions (as is relevant to the functional renormalization group programme). By this means, we are able to resolve an observation made in Ref.~\cite{Millington:2021ftp}, in the case of a zero-dimensional $\phi^4$ theory, that the flow equation for the coefficient of the term quartic in the would-be scalar field $\phi$ does not match the expected flow equation for the four-point vertex.

The remainder of this article is organised as follows.  In Sec.~\ref{sec:2PIEA}, we first review the definition of the 2PI effective action, its convexity, and the extraction of the two-point function, all in the context of a zero-dimensional, scalar quantum field theory. We then proceed in Sec.~\ref{sec:VertexFunctions} to describe a procedure for extracting $n$-point vertices from partial derivatives of the 2PI effective action, before applying it to the derivation of the flow equation for the four-point vertex at second-order in perturbation theory in Sec.~\ref{sec:RGFlows} within the approach of Refs.~\cite{Alexander:2019cgw, Millington:2021ftp}. The generalisation of these results to non-zero dimensions is presented in Sec.~\ref{sec:ToFieldTheory}. Our concluding remarks are provided in Sec.~\ref{sec:Conc}.


\section{2PI Effective Action}
\label{sec:2PIEA}

The two-particle-irreducible (2PI) effective action~\cite{Cornwall:1974vz} has the form
\begin{equation}
    \label{eq:2PIform}
    \Gamma(\phi,\Delta)=\mathcal{W}(\mathcal{J},\mathcal{K})+\mathcal{J}\phi+\frac{1}{2}\mathcal{K}(\phi^2+\hbar\Delta),
\end{equation}
wherein $\mathcal{W}(J,K)=-\hbar\ln\mathcal{Z}(J,K)$ and, for our zero-dimensional example, 
\begin{equation}
    \mathcal{Z}(J,K)=\mathcal{N}\int\ud\Phi\exp\left\{ -\frac{1}{\hbar}\left[ S(\Phi)-J\Phi-\frac{1}{2}K\Phi^2\right]\right\}
\end{equation}
is the path integral in the presence of sources $J$ and $K$ with normalization $\mathcal{N}$.  For definiteness, we take the classical action to have the form
\begin{equation}\label{eq:classical_action}
    S(\Phi)=\frac{1}{2!}\Phi^2+\frac{\lambda}{4!}\Phi^4,
\end{equation}
with $\lambda>0$.

The particular sources $\mathcal{J}\equiv\mathcal{J}(\phi,\Delta)$ and $\mathcal{K}\equiv \mathcal{K}(\phi,\Delta)$ are fixed by extremisation of the effective action such that they are functions of the ``one-'' and ``two-point'' variables $\phi$ and $\Delta$ (as emphasized in Refs.~\cite{Vasilev1, Vasilev2, Vasilev:1973vd, Garbrecht:2015cla}). The latter are given by
\begin{subequations}
\begin{align}
    \phi&=-\frac{\del}{\del \mathcal{J}}\mathcal{W}(\mathcal{J},\mathcal{K}),\\
    \hbar\Delta&=-2\frac{\del}{\del \mathcal{K}}\mathcal{W}(\mathcal{J},\mathcal{K})-\phi^2=-\hbar\frac{\del^2}{\del\mJ^2}\mW(\mJ,\mK).
\end{align}
\end{subequations}

The equations of motion (algebraic equations, in the zero-dimensional case) for $\phi$ and $\Delta$ are obtained by taking partial derivatives of the effective action:
\begin{subequations}
\begin{align}
       \frac{\del}{\del\phi}\Gamma(\phi,\Delta)&=\mJ(\phi,\Delta)+\mK(\phi,\Delta)\phi,\\\label{eq:Gamma_Delta}
    \frac{\del}{\del\Delta}\Gamma(\phi,\Delta)&=\frac{\hbar}{2}\mK(\phi,\Delta),
\end{align}
\end{subequations}
where partial derivatives with respect to $\phi$ are understood at fixed $\Delta$ and vice versa.

The 2PI effective action is a double Legendre transform of the Schwinger function $\mathcal{W}(J,K)$ with respect to the sources $J$ and $K$. By inspection of Eq.~\eqref{eq:2PIform}, we should anticipate that the convex-conjugate pairs of variables are
\begin{subequations}
\begin{align}
    \mJ'&=\mJ,\\
    \mK'&=\frac{1}{2}\mK,
\end{align}
\end{subequations}
and
\begin{subequations}
\label{eq:PrimedVariables}
\begin{align}
    \phi'(\phi,\Delta)&=\phi,\\
    \Delta'(\phi,\Delta)&=\phi^2+\hbar\Delta
\end{align}
\end{subequations}
(not $\phi$ and $\Delta$). These definitions lead to
\begin{subequations}
\begin{align}\label{eq:phi_prime_deriv}
    \frac{\del}{\del\phi'}&=\frac{\del}{\del\phi}-\frac{2}{\hbar}\phi\frac{\del}{\del\Delta},\\\label{eq:Delta_prime_deriv}
    \frac{\del}{\del\Delta'}&=\frac{1}{\hbar}\frac{\del}{\del\Delta},
\end{align}
\end{subequations}
and it can be proven that $\Gamma(\phi,\Delta)$ is a convex function of $(\phi',\Delta')$~\cite{Millington:2019nkw}. It is important to note that the 2PI effective action is not, in general, convex with respect to the pair $(\phi,\Delta)$~\cite{Millington:2019nkw}.

The convexity of the 2PI effective action leads to four identities, encoded in the product~\cite{Vasilev1, Vasilev2}
\begin{equation}
    \label{eq:identity}
    {\rm Hess}(-\mathcal{W}){\rm Hess}(\Gamma)=\mathbb{1},
\end{equation}
where the Hessian of $-\mathcal{W}$ is taken with respect to $\mathcal{J}'$ and $\mathcal{K}'$, and the Hessian of $\Gamma$ is taken with respect to $\phi'$ and $\Delta'$. These four identities, along with the definitions of the convex conjugate pairs of variables in Eq.~\eqref{eq:PrimedVariables}, allow us to express the partial derivatives of $\mathcal{W}$ with respect to the original sources $\mathcal{J}$ and $\mathcal{K}$ in terms of the derivatives of the 2PI effective action with respect to the variables $\phi$ and $\Delta$~\cite{Millington:2021ftp} (equivalent identities appear in footnote 11 of Ref.~\cite{Cornwall:1974vz}):
\begin{subequations}
\label{eq:convexity_identities}
\begin{align}
\label{eq:twopoint}
    \frac{\del^2\mW}{\del \mJ^2}&=
    -\Delta=-\left[\frac{\partial^2\Gamma}{\partial\phi^2}-\frac{2}{\hbar}\frac{\partial \Gamma}{\partial \Delta}-\left(\frac{\partial^2 \Gamma}{\partial \phi\partial\Delta}\right)\left(\frac{\partial^2\Gamma}{\partial\Delta^2}\right)^{-1}\left(\frac{\partial^2 \Gamma}{\partial \phi\partial\Delta}\right)\right]^{-1},\\
    \label{eq:W_JK}
    \frac{\del^2\mW}{\del\mJ\del\mK}&=\frac{\hbar}{2}\left(\frac{\del^2\Gamma}{\del\Delta^2} \right)^{-1}\left(\frac{\del^2\Gamma}{\del\phi\del\Delta}-\frac{2}{\hbar}\phi\frac{\del^2\Gamma}{\del\Delta^2} \right)\Delta,\\
    \frac{\del^2\mW}{\del\mK^2}&=-\frac{\hbar^2}{4}\left(\frac{\partial^2 \Gamma}{\partial \Delta^2}\right)^{-1}\Delta\left(\frac{\partial^2 \Gamma}{\partial \phi^2}-\frac{2}{\hbar}\frac{\partial \Gamma}{\partial \Delta}-\frac{4}{\hbar}\phi\frac{\partial^2\Gamma}{\partial\phi\partial\Delta}+\frac{4}{\hbar^2}\phi^2\frac{\partial^2\Gamma}{\partial \Delta^2}\right).
\end{align}
\end{subequations}


\section{Extracting $n$-Point Vertex Functions}
\label{sec:VertexFunctions}

By application of the chain rule, we can show that
\begin{equation}
    \begin{pmatrix} \frac{\partial}{\partial \mJ'} \\ \frac{\partial}{\partial \mK'} \end{pmatrix}=\begin{pmatrix} \frac{\partial \phi'}{\partial \mJ'} & \frac{\partial \Delta'}{\partial \mJ'}\\ \frac{\partial \phi'}{\partial \mK'} & \frac{\partial \Delta'}{\partial \mK'}\end{pmatrix}\begin{pmatrix} \frac{\partial}{\partial \phi'} \\ \frac{\partial}{\partial \Delta'} \end{pmatrix}={\rm Hess}(-\mathcal{W})\begin{pmatrix} \frac{\partial}{\partial \phi'} \\ \frac{\partial}{\partial \Delta'} \end{pmatrix}.
\end{equation}
Similarly,
\begin{equation}
    \begin{pmatrix} \frac{\partial}{\partial \phi'} \\ \frac{\partial}{\partial \Delta'}\end{pmatrix}={\rm Hess}(\Gamma)\begin{pmatrix} \frac{\partial}{\partial \mJ'} \\ \frac{\partial}{\partial \mK'} \end{pmatrix}.
\end{equation}
These expressions are mutually consistent by virtue of Eq.~\eqref{eq:identity}.
Introducing the coordinates $f^a\equiv(\phi',\Delta')$ and $f_a\equiv(\mathcal{J}',\mathcal{K}')$, and the derivatives
\begin{subequations}
\begin{align}
\partial_a&\equiv \frac{\partial}{\partial f^a}=\left(\frac{\partial}{\partial \phi'},\frac{\partial}{\partial \Delta'}\right),\\
\partial^a&\equiv \frac{\partial}{\partial f_a}=\left(\frac{\partial}{\partial \mathcal{J}'},\frac{\partial}{\partial \mathcal{K}'}\right),
\end{align}
\end{subequations}
we can write
\begin{subequations}
\begin{align}
    \partial_a&=M_{ab}\partial^b,\\
    \partial^a&=M^{ab}\partial_b,
\end{align}
\end{subequations}
where
\begin{subequations}
\begin{align}
M^{ab}&={\rm Hess}(-\mathcal{W})^{ab},\\
M_{ab}&={\rm Hess}(\Gamma)_{ab},
\end{align}
\end{subequations}
with [from Eq.~\eqref{eq:identity}]
\begin{equation}
    M_{ab}M^{bc}=\delta_a^c.
\end{equation}
Note that $M$ behaves like a metric on the configuration space and the dual space of convex-conjugate variables. This suggests an intriguing methodology for dealing with the functional identities arising from the $n{\rm PI}$ effective action that may be related to the geometry of Hessian manifolds (see, e.g., Ref.~\cite{ShimaYagi}) and information manifolds (see, e.g., Ref.~\cite{Nielsen}), and we leave further discussion of this to future work.

By the above means, and using Eqs.~\eqref{eq:phi_prime_deriv} and \eqref{eq:Delta_prime_deriv}, we can express the derivative with respect to $\mJ'$ in terms of derivatives with respect to $\phi$ and $\Delta$:
\begin{align}
\label{eq:Jderivative}
\frac{\partial}{\partial \mJ}\equiv \frac{\partial}{\partial \mJ'}&= \frac{\del\phi'}{\del\mJ'}\frac{\del}{\del\phi'}+\frac{\del\Delta'}{\del\mJ'}\frac{\del}{\del\Delta'}\nonumber\\
&=-\frac{\partial^2\mathcal{W}}{\partial \mJ^2}\frac{\partial}{\partial \phi'}-2\frac{\partial^2\mathcal{W}}{\partial \mathcal{J}\partial \mathcal{K}}\frac{\partial}{\partial \Delta'}\nonumber\\
&=\Delta\left( \frac{\del}{\del\phi}-\frac{2}{\hbar}\phi\frac{\del}{\del\Delta}\right)-2\left[ \frac{\hbar}{2}\left(\frac{\del^2\Gamma}{\del\Delta^2}\right)^{-1}\left( \frac{\del^2\Gamma}{\del\phi\del\Delta}-\frac{2}{\hbar}\phi\frac{\del^2\Gamma}{\del\Delta^2}\right)\Delta\right]\frac{1}{\hbar}\frac{\del}{\del\Delta}\nonumber\\
&=\Delta\left[\frac{\partial}{\partial \phi}-\frac{\partial^2\Gamma}{\partial\phi\partial\Delta}\left(\frac{\partial^2\Gamma}{\partial\Delta^2}\right)^{-1}\frac{\partial}{\partial \Delta}\right].
\end{align}
An equivalent operator first appeared in Ref.~\cite{Vasilev1}.

The two-point function is given in Eq.~\eqref{eq:twopoint}, and the various $n$-point functions can then be obtained by taking $n-2$ derivatives of the two-point function with respect to $\mathcal{J}$ by repeated application of Eq.~\eqref{eq:Jderivative}, cf.~the equivalent approach of Ref.~\cite{Vasilev1}, before multiplying by $(\hbar\Delta)^{-n}$ to amputate the external two-point functions. Specifically, the connected $n$-point function (for $n>2$) is
\begin{equation}
    \braket{\phi^n}_{\rm conn}=\left( \hbar\frac{\del}{\del \mJ}\right)^{n-2}\left( -\hbar\frac{\del^2W}{\del \mJ^2}\right)=\left( \hbar\frac{\del}{\del \mJ}\right)^{n-2}(\hbar\Delta),
\end{equation}
and the amputated $n$-point vertex can be written as
\begin{equation}
\label{eq:vertices}
\Gamma^{(n>2)}=-\hbar(\hbar\Delta)^{-n}\braket{\phi^n}_{\rm conn}=-\Delta^{-n}\left\{\Delta\left[\frac{\partial}{\partial \phi}-\frac{\partial^2\Gamma}{\partial\phi\partial\Delta}\left(\frac{\partial^2\Gamma}{\partial\Delta^2}\right)^{-1}\frac{\partial}{\partial \Delta}\right]\right\}^{n-2} \Delta.
\end{equation}
The overall minus sign and factor of $\hbar$ are such that $\Gamma^{(n>2)}$ coincides with the tree-level vertex to zeroth order in $\hbar$; e.g., for the action in Eq.~\eqref{eq:classical_action},
\begin{equation}
\Gamma^{(4)}=\lambda+\mathcal{O}(\hbar).
\end{equation}

To order $\lambda^2$, the explicit expression for the 2PI effective is (see Refs.~\cite{Millington:2019nkw, Millington:2021ftp} for the complete derivation)
\begin{align}
    \Gamma(\phi,\Delta)&=\frac{1}{2}\phi^2+\frac{\lambda}{4!}\phi^4+\frac{\hbar}{2}\left[\ln\Delta^{-1}+\left(1+\frac{\lambda}{2}\phi^2\right)\Delta -1\right]\nonumber\\&\phantom{=}+\hbar^2\left[\frac{\lambda}{8}\Delta^2-\frac{\lambda^2}{12}\phi^2\Delta^3\right]-\frac{\hbar^3\lambda^2}{48}\Delta^4.
\end{align}
Hence,
\begin{subequations}
\begin{align}
    \frac{\partial^2\Gamma}{\partial\phi\partial\Delta}=\frac{\hbar\lambda}{2}\phi+\mathcal{O}(\lambda^2),\\
    \frac{\partial^2\Gamma}{\partial\Delta^2}=\frac{\hbar}{2}\Delta^{-2}+\mathcal{O}(\lambda),
\end{align}
\end{subequations}
and so
\begin{align}
    \Gamma^{(4)}=\Delta^{-3}\left[\frac{\partial}{\partial \phi}-\lambda\phi\Delta^2\frac{\partial}{\partial \Delta}\right]\lambda\phi\Delta^3=\lambda-3\lambda^2\phi^2\Delta+\mathcal{O}(\lambda^3).
\end{align}
We emphasise that $\phi$ and $\Delta$ are independent variables, such that $\partial\Delta/\partial\phi=0$ and $\partial\phi/\partial\Delta=0$. Notice that all of these expressions are obtained directly from the 2PI effective action in terms of the variables $\phi$ and $\Delta$.

For comparison, the equivalent procedure for the 1PI effective action involves taking only derivatives with respect to $\phi$, i.e.,
\begin{equation}
\Gamma^{(n>2)}_{\rm 1PI}=-\Delta^{-n}\left\{\Delta\frac{\partial}{\partial\phi}\right\}^{n-2}\Delta,
\end{equation}
where
\begin{equation}
    \Delta^{-1}=\frac{\partial^2\Gamma_{\rm 1PI}}{\partial \phi^2}.
\end{equation}
Thus, the $4$-point vertex has the simple expression
\begin{equation}
    \Gamma^{(4)}_{\rm 1PI}=\frac{\partial^4\Gamma_{\rm 1PI}}{\partial \phi^4}-3\Gamma^{(3)}_{\rm 1PI}\Delta\Gamma^{(3)}_{\rm 1PI}.
\end{equation}
We stress that the variables $\phi$ and $\Delta$ are not independent in the case of the 1PI effective action. In the 2PI case, the additional degree of freedom provided by $\mathcal{K}$ ensures that the variables $\phi$ and $\Delta$ are independent.


\section{Vertex RG Flows}
\label{sec:RGFlows}

We now turn our attention to the flow equations for the 2PI effective action of our zero-dimensional theory. The main aim of this section (and indeed this work) is to go beyond Ref.~\cite{Millington:2021ftp} and to compare the flow equations for the quartic vertex, as obtained from the usual 1PI framework and the 2PI approach of Refs.~\cite{Alexander:2019cgw, Millington:2021ftp}, the latter of which we now describe.

We proceed by promoting the source $\mathcal{K}\to \mathcal{R}_k$, which plays the role of the regulator in the 2PI approach for deriving exact flow equations with the parameter $k$ emulating the RG scale.\footnote{We use a non-standard sign convention in the definition of the regulator $\mathcal{R}_k$.} The variation of the 2PI effective action with respect to the scale $k$ is then
\begin{equation}
    \partial_k\Gamma(\phi,\Delta_k)=\frac{\hbar}{2}\mathcal{R}_k\partial_k\Delta_k,
\end{equation}
where $\Delta_k$ is given by Eq.~\eqref{eq:twopoint} (see Refs.~\cite{Alexander:2019cgw, Millington:2021ftp}).

In order to derive the flow equations, we make the following Ansatz for the 2PI effective action:
\begin{equation}\label{eq:eff_action_ansatz}
    \Gamma(\phi,\Delta_k)=\alpha_k(\Delta_k)+\frac{1}{2}\beta_k(\Delta_k)\phi^2+\frac{1}{4!}\gamma_k(\Delta_k)\phi^4.
\end{equation}
Using (\ref{eq:twopoint}) and (\ref{eq:Gamma_Delta}), along with $\mathcal{K}\to\mathcal{R}_k$, this leads to the following expression for the inverse two-point function~\cite{Millington:2021ftp}:
\begin{equation}\label{eq:Delta_inv}
    \Delta^{-1}_k=\beta_k-\mathcal{R}_k(\phi,\Delta_k)+\frac{1}{2}\gamma_k\phi^2-\left[\phi\frac{\partial\beta_k}{\partial \Delta_k}+\frac{1}{3!}\phi^3\frac{\partial\gamma_k}{\partial \Delta_k}\right]^2\left[\frac{\partial^2\alpha_k}{\partial \Delta_k^2}+\frac{1}{2}\phi^2\frac{\partial^2\beta_k}{\partial \Delta_k^2}+\frac{1}{4!}\phi^4\frac{\partial^2\gamma_k}{\partial \Delta_k^2}\right]^{-1},
\end{equation}
where
\begin{equation}
    \mathcal{R}_k(\phi,\Delta_k)=\frac{2}{\hbar}\left[\frac{\partial\alpha_k}{\partial \Delta_k}+\frac{1}{2}\phi^2\frac{\partial\beta_k}{\partial \Delta_k}+\frac{1}{4!}\phi^4\frac{\partial\gamma_k}{\partial \Delta_k}\right]
\end{equation}
and we have suppressed the arguments of the $\alpha_k$, $\beta_k$ and $\gamma_k$ for conciseness. Their flow equations can be extracted by taking derivatives of the flow equation with respect to $\phi$ and $\Delta$ and evaluating at $\phi=0$, as described in Ref.~\cite{Millington:2021ftp}.

If we intend to work to order $\lambda^2$, this procedure leads to the following system of equations for the $\Delta$-derivatives of $\alpha(\Delta)$, $\beta(\Delta)$ and $\gamma(\Delta)$~\cite{Millington:2021ftp}:
\begin{subequations}
\label{eq:Deltaderivs}
\begin{align}
    \frac{\partial \alpha_k(\Delta)}{\partial \Delta}&=\frac{\hbar}{2}\mathcal{R}_k(0,\Delta),\\
    \frac{\partial^2 \alpha_k(\Delta)}{\partial \Delta^2}&=\frac{\hbar}{2}\left[\beta_k(\Delta)-\mathcal{R}_k(0,\Delta)\right]^2+\frac{\hbar^2}{4}\gamma_k(\Delta)-\frac{\hbar^3}{4}\gamma^2_k(\Delta)\left[\beta_k(\Delta)-\mathcal{R}_k(0,\Delta)\right]^{-2},\\
    \frac{\partial^3 \alpha_k(\Delta)}{\partial \Delta^3}&=-\hbar \left[\beta_k(\Delta)-\mathcal{R}_k(0,\Delta)\right]^3-\frac{\hbar^3}{2}\gamma_k^2(\Delta)\left[\beta_k(\Delta)-\mathcal{R}_k(0,\Delta)\right]^{-1},\\
    \frac{\partial \beta_k(\Delta)}{\partial \Delta}&=\frac{\hbar}{2}\gamma_k(\Delta)-\frac{\hbar^2}{2}\gamma^2_k(\Delta)\left[\beta_k(\Delta)-\mathcal{R}_k(0,\Delta)\right]^{-2},\\
    \frac{\partial^2 \beta_k(\Delta)}{\partial \Delta^2}&=-\hbar^2\gamma^2_k(\Delta)\left[\beta_k(\Delta)-\mathcal{R}_k(0,\Delta)\right]^{-1},\\
    \frac{\partial \gamma_k(\Delta)}{\partial \Delta}&=\mathcal{O}(\gamma_k^4).
    \label{eq:gamma_running}
\end{align}
\end{subequations}
Note that the regulator is evaluated at $\phi=0$ in the above expressions.

We see that the ``coupling parameter'' $\gamma_k(\Delta)$ is fixed straightforwardly by our boundary condition, i.e., $\gamma_k(\Delta)=\lambda$, at this order. A potentially striking observation (as made in Ref.~\cite{Millington:2021ftp}) is that $\gamma_k$ does not run until order $\lambda^4$, when, under the assumption that $\gamma_k$ is related to the four-point vertex, we might expect it to run at order $\lambda^2$.

The remaining flow equations are~\cite{Millington:2021ftp}
\begin{subequations}
\begin{gather}
    \partial_k\alpha_k(\Delta_k)=\frac{\hbar}{2}\mathcal{R}_k(0,\Delta_k)\partial_k\left[\beta_k(\Delta)-\mathcal{R}_k(0,\Delta)\right]^{-1},\\
    \partial_k\beta_k(\Delta_k)=\left\{\frac{\hbar}{2}\lambda-\frac{\hbar^2}{2}\lambda^2\left[\beta_k(\Delta)-\mathcal{R}_k(0,\Delta)\right]^{-2}\right\}\partial_k\left[\beta_k(\Delta)-\mathcal{R}_k(0,\Delta)\right]^{-1}.
\end{gather}
\end{subequations}
The solutions to order $\lambda^2$ are~\cite{Millington:2021ftp}
\begin{subequations}
\begin{align}    
    \beta_k(\Delta_k)&=\beta_0+\frac{\hbar \lambda}{2}\frac{1}{1-\mathcal{R}_k(0,\Delta_k)}-\frac{5\hbar^2\lambda^2}{12}\frac{1}{[1-\mathcal{R}_k(0,\Delta_k)]^3},\\
    \alpha_k(\Delta_k)&=\alpha_0+\frac{\hbar}{2}\frac{1}{1-\mathcal{R}_k(0,\Delta_k)}+\frac{\hbar}{2}\ln[1-\mathcal{R}_k(0,\Delta)]+\frac{\hbar^2\lambda}{8}\frac{1-3\mathcal{R}_k(0,\Delta_k)}{[1-\mathcal{R}_k(0,\Delta)]^3}\nonumber\\&\phantom{=}-\frac{\hbar^3\lambda^2}{12}\frac{1-5\mathcal{R}_k(0,\Delta_k)}{[1-\mathcal{R}_k(0,\Delta_k)]^5},
\end{align}
\end{subequations}
where $\alpha_0$ and $\beta_0$ are integration constants. We fix $\beta_0=1$ by matching to the limit $\lambda \to 0$.

At this point, we are, in some sense, finished, since we have all that we need to reconstruct the 2PI effective action and all of the resulting $n$-point functions, correct to order $\lambda^2$, as was done in Ref.~\cite{Millington:2021ftp}. However, it is instructive to obtain the four-point vertex and derive its flow equation, both evaluated at $\phi=0$, as we will now do. In particular, this will allow us to understand why the coefficient $\gamma_k$ of the $\phi^4$ term in the Ansatz~\eqref{eq:eff_action_ansatz} does not appear to run at order $\lambda^2$ [see Eq.~\eqref{eq:gamma_running} above].

We first use Eq.~\eqref{eq:vertices} with $n=3$ to calculate the expression for the three-point vertex $\Gamma^{(3)}$ of the effective action in Eq.~\eqref{eq:eff_action_ansatz}, which we need only to first order in $\phi$ (since we will subsequently take one more derivative with respect to $\phi$, before setting $\phi=0$):
\begin{align}
    \Gamma^{(3)}_k&=\Delta_k^{-2}\frac{\del^2\Gamma_k}{\del\phi\del\Delta_k}\left( \frac{\del^2\Gamma_k}{\del\Delta^2_k}\right)^{-1}\nonumber\\
    &=\Delta^{-2}_k\left[ \phi\frac{\partial\beta_k}{\partial \Delta_k}+\frac{\phi^3}{3!}\frac{\partial \gamma_k}{\partial\Delta_k}\right]\left[\frac{\partial^2\alpha_k}{\partial\Delta_k^2}+\frac{1}{2}\phi^2\frac{\partial^2\beta_k}{\partial\Delta_k^2}+\frac{1}{4!}\phi^4\frac{\partial^2\gamma_k}{\partial\Delta^2_k} \right]^{-1}\nonumber\\
    &=\Delta^{-2}_k\phi\frac{\partial\beta_k}{\partial \Delta_k}\left(\frac{\partial^2\alpha_k}{\partial\Delta^2_k}\right)^{-1}+\mathcal{O}(\phi^3).
\end{align}
Multiplying this result by $(\hbar\Delta_k)^3$, differentiating with respect to $\mathcal{J}/\hbar$ via Eq.~\eqref{eq:Jderivative}, and amputating four factors of $\hbar\Delta_k$, only the $\phi$ derivative contributes at $\phi=0$, and we obtain
\begin{equation}
    \Gamma^{(4)}_k|_{\phi=0}=\left.\Delta_k^{-2}\frac{\partial\beta_k}{\partial \Delta_k}\left(\frac{\partial^2\alpha_k}{\partial\Delta^2_k}\right)^{-1}\right|_{\phi=0}.
\end{equation}
We need only the lowest-order terms in the expressions for $\frac{\partial \beta_k(\Delta_k)}{\partial \Delta_k}$ and $\frac{\partial^2 \alpha_k(\Delta_k)}{\partial \Delta_k^2}$ from Eq.~\eqref{eq:Deltaderivs} along with $\Delta_k$ from Eq.~\eqref{eq:Delta_inv}, giving
\begin{equation}
    \Gamma^{(4)}_k|_{\phi=0}=\gamma_k(\Delta_k)-\frac{3}{2}\hbar\gamma^2_k(\Delta_k)\frac{1}{[\beta_k(\Delta_k)-\mathcal{R}_k(0,\Delta_k)]^2}.
\end{equation}

Since the coupling parameter $\gamma_k(\Delta_k)$ does not flow until order $\lambda^4$ in the 2PI approach~\cite{Millington:2021ftp}, the flow equation for $\Gamma^{(4)}_k|_{\phi=0}$ to order $\lambda^2$ is
\begin{equation}
    \label{eq:Gamma4flow}
    \partial_k\Gamma^{(4)}_k|_{\phi=0}=-3\hbar\lambda^2\frac{\partial_k\mathcal{R}_k(0,\Delta_k)}{[1-\mathcal{R}_k(0,\Delta_k)]^3},
\end{equation}
wherein we have used the fact that $\beta_k(\Delta_k)=1+\mathcal{O}(\gamma_k)$ and  $\partial_k\beta_k(\Delta_k)$ is order $\gamma_k(\Delta_k)$. Most significantly, this agrees with the flow equation for the four-point vertex of the average 1PI approach, also at order $\lambda^2$, as we describe below.

The average 1PI effective action is given by~\cite{Wetterich:1989xg} (for a review, see, e.g., Ref.~\cite{Berges:2000ew})
\begin{equation}
    \Gamma^{\rm av}_{\rm 1PI}(\phi,\mathcal{R}_k)=\mathcal{W}(\mathcal{J},\mathcal{R}_k)+\mathcal{J}(\phi)\phi+\frac{1}{2}\mathcal{R}_k\phi^2.
\end{equation}
It differs from the 2PI effective action in that there has been no Legendre transform with respect to the source $\mathcal{R}_k$. As a result, its natural variables are $\phi$ and $\mathcal{R}_k$, and it is for this reason that $\phi$ and $\Delta$ are {\it not} independent variables in the 1PI case. The flow equation is~\cite{Wetterich:1992yh, Morris:1993qb, Ellwanger:1993mw}
\begin{equation}
\partial_k\Gamma^{\rm av}_{\rm 1PI}=-\frac{\hbar}{2}\Delta_k\partial_k\mathcal{R}_k,
\end{equation}
and we draw attention to our non-standard choice of sign for the definition of the regulator. The two-point function is given in terms of derivatives of $\Gamma^{\rm av}_{\rm 1PI}$ via
\begin{equation}
    \Delta^{-1}_k=\frac{\partial^2\Gamma^{\rm av}_{\rm 1PI}}{\partial \phi^2}-\mathcal{R}_k,
\end{equation}
cf.~Eq.~\eqref{eq:twopoint}, wherein the additional terms illustrate the difference in the degree of resummation provided by the 2PI approach.

Proceeding by making the Ansatz
\begin{equation}
    \label{eq:1PIAnsatz}
    \Gamma^{\rm av}_{\rm 1PI}(\phi,\mathcal{R}_k)=\tilde{\alpha}_k(\mathcal{R}_k)+\frac{1}{2}\tilde{\beta}_k(\mathcal{R}_k)\phi^2+\frac{1}{4!}\tilde{\gamma}_k(\mathcal{R}_k)\phi^4,
\end{equation}
the flow equations are obtained by taking $\phi$ derivatives at $\phi=0$. Working to order $\lambda^2$, we obtain~\cite{Millington:2021ftp}
\begin{subequations}
\begin{align}
\partial_k\tilde{\alpha}_k&=-\frac{\hbar}{2}\frac{\partial_k\mathcal{R}_k}{\tilde{\beta}_k-\mathcal{R}_k},\\
\partial_k\tilde{\beta}_k&=\frac{\hbar}{2}\frac{\tilde{\gamma}_k\partial_k\mathcal{R}_k}{[\tilde{\beta}_k-\mathcal{R}_k]^2},\\
\label{eq:tildegammaflow}
\partial_k\tilde{\gamma}_k&=-3\hbar\frac{\tilde{\gamma}_k^2\partial_k\mathcal{R}_k}{[\tilde{\beta}_k-\mathcal{R}_k]^3}.
\end{align}
\end{subequations}
The solutions strictly to order $\lambda^2$ are~\cite{Millington:2021ftp}
\begin{subequations}
\begin{align}
    \tilde{\alpha}_k&=\tilde{\alpha}_0+\frac{\hbar}{2}\ln(1-\mathcal{R}_k)+\frac{\hbar^2\lambda}{8}\frac{1}{(1-\mathcal{R}_k)^2}-\frac{\hbar^3\lambda^2}{12}\frac{1}{(1-\mathcal{R}_k)^4},\\
    \tilde{\beta}_k&=1+\frac{\hbar\lambda}{2}\frac{1}{1-\mathcal{R}_k}-\frac{5\hbar^2\lambda^2}{12}\frac{1}{(1-\mathcal{R}_k)^3},\\
    \tilde{\gamma}_k&=\lambda-\frac{3\hbar\lambda^2}{2}\frac{1}{(1-\mathcal{R}_k)^2}.
\end{align}
\end{subequations}

Notice that, in the average 1PI approach, the four-point vertex function at $\phi=0$ coincides with the coefficient $\tilde{\gamma}_k(\mathcal{R}_k)$ of $\phi^4$ in the Ansatz \eqref{eq:1PIAnsatz} for the average 1PI effective action, i.e., $\Gamma^{{\rm av}(4)}_{\rm 1PI}|_{\phi=0}=\tilde{\gamma}_k(\mathcal{R}_k)$. This is not so for the 2PI case, because, in the 2PI resummation, the $\lambda^2$ correction to the four-point vertex is absorbed in $\alpha_k(\Delta_k)$ and $\beta_k(\Delta_k)$, and not in the coefficient $\gamma_k(\Delta_k)$ of $\phi^4$. It is for this reason that the coupling parameter $\gamma_k(\Delta_k)$ does not run at order $\lambda^2$, whereas $\tilde{\gamma}_k(\mathcal{R}_k)$ of the 1PI approach does. With these observations, and comparing Eqs.~\eqref{eq:Gamma4flow} and \eqref{eq:tildegammaflow}, we see that the flows of the four-point vertex at $\phi=0$, as obtained from the 2PI and average 1PI approaches, agree at order $\lambda^2$. Due to the differing resummation implicit in the 2PI versus 1PI approaches, however, there is no reason why these flow equations should agree once we move away from the strictly fixed-order result.


\section{Field theory generalisation}
\label{sec:ToFieldTheory}

In order to extend the previous results to non-zero dimensions, it is convenient to consider first the generalisation to $N$ fields, i.e., $\phi\to \phi^{\alpha}$ and $\Delta\to\Delta^{\alpha\beta}$ with $\alpha,\beta=1,2,\dots,N$. In the latter case, the 2PI effective action becomes
\begin{equation}
    \label{eq:2PImultifield}
    \Gamma(\{\phi\},\{\Delta\})=\mathcal{W}(\{\mathcal{J}\},\{\mathcal{K}\})+\mathcal{J}_{\alpha}\phi^{\alpha}+\frac{1}{2}\mathcal{K}_{\alpha\beta}(\phi^{\alpha}\phi^{\beta}+\hbar\Delta^{\alpha\beta}).
\end{equation}

A comprehensive discussion of how the convexity properties can be translated to the multifield case, along with the resulting identities (cf.~Eq.~\eqref{eq:convexity_identities}), is provided in Ref.~\cite{Millington:2021ftp}. Making use of those results, the derivative with respect to the source $\mathcal{J}_{\alpha}$, generalising Eq.~\eqref{eq:Jderivative}, becomes
\begin{equation}
    \frac{\del}{\del \mJ_{\alpha}}=\Delta^{\alpha\beta}\left[ \frac{\del}{\del\phi^{\beta}}-\frac{\del^2\Gamma}{\del\phi^{\beta}\del\Delta^{\gamma\delta}}\left( \frac{\del^2\Gamma}{\del\Delta^{\gamma\delta}\del\Delta^{\rho\sigma}}\right)^{-1}\frac{\del}{\del\Delta^{\rho\sigma}}\right].
\end{equation}
The expression for the $n$-point vertex [Eq.~\eqref{eq:vertices}] is then promoted to
\begin{align}
\Gamma^{(n>2)}_{\alpha_1\alpha_2\dots\alpha_n}&=-\hbar^{n+1}\Delta^{-1}_{\alpha_1\beta_1}\Delta^{-1}_{\alpha_2\beta_2}\cdots\Delta^{-1}_{\alpha_{n}\beta_{n}}\braket{\phi^{\beta_1}\phi^{\beta_2}\cdots\phi^{\beta_n}}_{\rm conn}\nonumber\\ &=-\Delta^{-1}_{\alpha_1\beta_1}\Delta^{-1}_{\alpha_2\beta_2}\cdots\Delta^{-1}_{\alpha_{n}\beta_{n}}\frac{\partial}{\partial \mathcal{J}_{\beta_1}}\frac{\partial}{\partial \mathcal{J}_{\beta_2}}\dots\frac{\partial}{\partial \mathcal{J}_{\beta_{n-2}}}\Delta^{\beta_{n-1}\beta_n}.
\end{align}

The generalisation from this multi-field case to the full field theory case in $d>0$ dimensions is straightforward:~we must simply interpret the field-space indices to include coordinate variables, which, when contracted, are integrated over with an appropriate spacetime measure, as per the conventions of the DeWitt notation.


\section{Concluding Remarks}
\label{sec:Conc}

We have outlined a procedure for extracting $n$-point vertex functions from the two-particle irreducible (2PI) effective action that exploits its convexity in the two-dimensional configuration space specified by the one and non-connected two-point functions.  This result is expected to increase the utility of approaches based on the 2PI effective action, including alternative derivations (see Refs.~\cite{Alexander:2019cgw, Millington:2021ftp}) of the exact flow equations for interacting quantum field theories that are complementary to long-standing approaches thanks to the differing ways that the resummation of loop corrections is organised in the 2PI versus 1PI frameworks. An explicit application of these results to the RG evolution of an interacting scalar field theory in $d>0$ spacetime dimensions will be provided in a revised version of a previous unpublished work~\cite{Alexander:2019quf}.


\begin{acknowledgments}

The Authors thank the organisers and participants of the 11\textsuperscript{th} International Conference on the Exact Renormalization Group 2022 (ERG2022) and especially Urko Reinosa for helpful discussions. The Authors also thank the referee of this work for introducing us to the early and insightful works of Vasil'ev~\cite{Vasilev:1973vd}, and Vasil'ev and Kazanskii~\cite{Vasilev1, Vasilev2}. This work was supported by a United Kingdom Research and Innovation (UKRI) Future Leaders Fellowship [Grant No.~MR/V021974/2] and by the Science and Technology Facilities Council (STFC) [Grant No.~ST/P000703/1 and ST/T000732/1]. For the purpose of open access, the authors have applied a Creative Commons Attribution (CC BY) licence to any Author Accepted Manuscript version arising.

\end{acknowledgments}


\section*{Data access statement}

No data were created or analysed in this study.


\end{document}